\RequirePackage{fix-cm}
\documentclass[smallextended]{svjour3}
\smartqed  
\usepackage[bookmarks = true
, pdfpagemode = None
, pdfstartview = FitH
, colorlinks = true
, urlcolor = blue]{hyperref}
\usepackage[braket, qm]{qcircuit}
\usepackage{graphicx,amsmath}
\usepackage{rotating}
\usepackage{multirow}
\usepackage{amsmath,amssymb,amsfonts,amsxtra,dsfont}
\usepackage{braket}
\usepackage{array}      
\usepackage{multirow}
\usepackage{booktabs}
\usepackage{relsize} 
                        
\newcommand\dsum{\displaystyle\sum}

\newcommand{\eins}{ {\rm 1}\mkern-4.5mu {\rm l}}
\newcommand{\bigO}{\mathcal{O}}
\newcommand{\gap}[1][=]{\mathrel{\phantom{#1}}} 

\begin{document}
\nocite{LesovikEtal2010,SuslovEtal2011,Bennett2014,Deutsch1992,Bernstein1997,Simon1994,Shor1994,Grover1996,Feynman1982,VanMeyer2013,Nielsen2000,Cleve1998,Brassard2002}
\title{A Quantum Abacus based encoding system
}


\author{J. V. \'Alvarez-Bravo\and J. J. \'Alvarez-S\'anchez\and I. Aparicio-Morgado}


\institute{J. V. \'Alvarez--Bravo \at
              Escuela de Ingenier\'ia Inform\'atica de Segovia \\
              Tel.: +34-921-112454\\
              Fax: +34-921-112400\\
              \email{jvalvarez@infor.uva.es}           
           \and
           J. J. \'Alvarez--S\'anchez \at
                         Escuela de Ingenier\'ia Inform\'atica de Segovia \\
                         Tel.: +34-921-112430\\
                         Fax: +34-921-112400\\
                         \email{jjalvarez@infor.uva.es}
            \and
            JI. Aparicio--Morgado \at
                                     Escuela de Ingenier\'ia Inform\'atica de Segovia \\
                                     Tel.: +34-921-112419\\
                                     Fax: +34-921-112400\\
                                     \email{ignacio@infor.uva.es}
}

\date{Received: date / Accepted: date}

\maketitle

\begin{abstract}
A formal description of a quantum abacus based encoding system is presented. This way of representing data for processing purposes is based on a quantum algorithm for counting qubits introduced by Lesovik et al. \cite{LesovikEtal2010} and Suslov et al. \cite{SuslovEtal2011}, but formally developed in this work. Finally, in order to illustrate the potential of this proposal, the implementation of the basic quantum array operations through this encoding system is presented.
\end{abstract}

\section{Introduction}
\label{intro}
Quantum computing paradigm is currently one of the most active and promising research area. From the first contributions, in areas such as cryptography \cite{Bennett2014} or speeding-up some classical algorithms \cite{Deutsch1992,Bernstein1997,Simon1994,Shor1994,Grover1996}, up to present, many works have laid the foundations about how to represent and process data under the rules of quantum mechanics. There are many reasons for this increasing interest, some of them were pointed out by Feynmann in his well-known dissertation \cite{Feynman1982},  but all of them can be summarized just in one: the Quantum computer is the only way of overcoming the limitations imposed to the Classical one by Moore's Law.  Differents approaches have been proposed for carrying out this great challenge \cite{VanMeyer2013}. The most accepted, the standard model, defines the quantum computer as a circuit-based computing system where a sequence of unitary gates act coherently on an input register of qubits \cite{Nielsen2000}. The aim of this work is to contribute from this last perspective providing an efficient quantum circuit for encoding and processing quantum data. The quantum encoding system presented in this work is based on the Quantum Abacus (QA), a quantum algorithm for counting qubits introduced by Lesovik et al. \cite{LesovikEtal2010} and Suslov et al. \cite{SuslovEtal2011} but formally described in our work  in terms of a deterministic QFT based Phase Estimation Algorithm. The main idea can be merely described as follows: since the output of the QA circuit is the number of qubits in a sequence, then the underlying representation can be used for  encoding any integer number. In this case, the underlying representation is a set of phase shifts standing for the QFT of the output. The key point is how the QA circuit allows to infer easely a compact expression for these phase shifts in terms of elementary quantum gates, the encoding circuit, and how this one can be used, for instance,  to implement the basic operations of a simple quantum data structure.  

The paper has been organized as follows: In section 2 a deterministic Phase Estimation Algorithm schema (PEA) based on QFT is described. In this regard, the zero failure rate condition is achieved and  used for implementing the case when the number of input qubits  is equal to the number of ancillary qubits. In section 3, the Quantum Abacus circuit is formally described in terms of a PEA schema. In section 4, the framework for quantum encoding, by means of a QA, is presented. In section 5, in order to shed light on how to use this proposal, the implementation of creating and updating Quantum array operation are presented. Finally, in section 6,  a brief set of conclusions summarize the contributions of this paper.  
\section{The QFT based Phase Estimation Algorithm: the zero failure rate condition}\label{S3:QFTBasedPhase}
\newcommand{\exptwo}[1][n]{2^#1 -1}
\newcommand{\Normqv}[1][1]{\dfrac{#1}{\sqrt{2^n}}}
\newcommand{\Expqu}[1][]{e^{#1 i \Normqv[2 \pi j k]}}
\newcommand{\SumExp}[3][]{\dsum_{#2}^{#3} x_j \Expqu[#1] } 

Let us suppose a $2^n \times 2^n$ unitary operator $U$ defined in terms of the problem to be solved. If $\ket{\psi_j}$ and $\lambda_j$ are eigenvectors and eigenvalues of $U$ respectively in such a way that:

\begin{equation}
U=\dsum_{j=0}^{2^n-1} \lambda_j \ket{\psi_j} \bra{\psi_j} \quad \text{where} \quad \lambda_j = e^{i 2 \pi \phi_j}\qquad \text{and} \quad \phi_j \in \left[ 0, 1 \right)
\end{equation}

\begin{figure}[h]
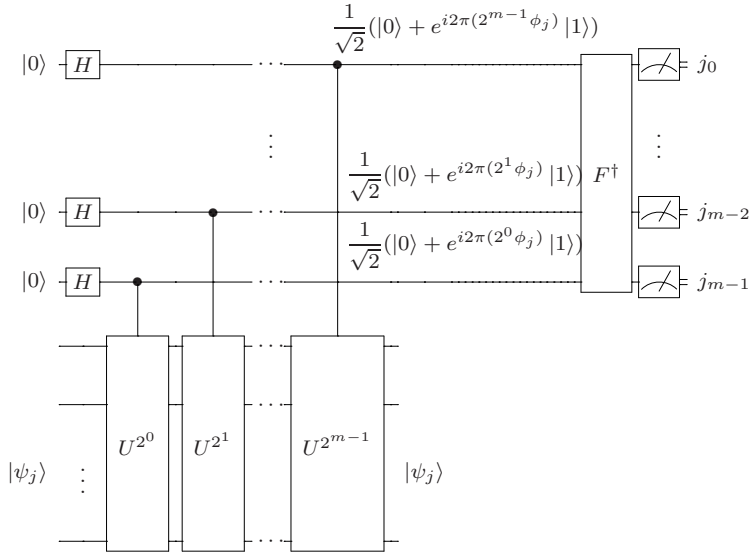

\include{2015_Figure03}
\caption{Detailed schema of QFT based PEA}\label{S3:Fig:SchemaQFT}
\end{figure}

The Phase Estimation Algorithm (PEA) is defined as the process of estimating $\phi_j$ with a specific precision when the input is a register containing the corresponding eigenvector $\ket{\psi_j}$ of $U$. In this work, a formulation of the PEA based on QFT will be considered \cite{Cleve1998} and is shown in Figure \ref{S3:Fig:SchemaQFT}.

Under this schema, the estimation process is efficiently carried out according to the following steps:
\begin{enumerate}
\item The ancillary register is prepared containing an evenly distributed superposition.
	The number of qubits of the ancillary register determines the estimation 
	accuracy.
\item Given a particular eigenvector $\ket{\psi_j}$ of $U$ as an input, the next step 
consists on kicking
	the corresponding phases in front of the ancillary register through the action 
	of the controlled powers
	of $U$ denoted as $c\_U^{2^l}$ for $l=m-1,\ldots ,0$, being $m$ the number of 
	qubits of the ancillary register.
\item In order to make $\phi_j$ an observable quantity in the computational basis, the 
inverse of QFT
	is applied on the ancillary register.
\item Then, a measurement process determines $\phi_j$ with a high probability 
and a precision lower than
	$1/2^m$.
\end{enumerate}

Adding up steps 1 to 4, a summarized formal description is presented:

\begin{gather}
\ket{0^{\otimes (m)}, \psi_j} \xrightarrow{H^{\otimes m} \otimes\hat{I}^{\otimes 
(n)}}
	\dfrac{1}{\sqrt{2^m}} \bigotimes_{l=m-1}^{0} \left( \ket{0} \ket{\psi_j} + \ket{1} 
	\ket{\psi_j} \right)	
	\xrightarrow{\otimes\hat{I}^{\otimes (m)} \otimes c\_U^{2^l}} \\
= \dfrac{1}{\sqrt{2^m}} \bigotimes_{l=m-1}^{0} 
	\left( \ket{0} \ket{\psi_j} + \ket{1} e^{i 2 \pi (2^l \phi_j) } \ket{\psi_j} \right) =
	\dfrac{1}{\sqrt{2^m}} \bigotimes_{l=m-1}^{0}
	\left( \ket{0} + e^{i 2 \pi (2^l \phi_j) } \ket{1} \right)  \ket{\psi_j} 
	\label{S3:eq:FormalAlgo}
\end{gather}

As can be observed in equation (\ref{S3:eq:FormalAlgo}), the phase shift is kicked 
in front of the ancillary register
and the eigenvector $\ket{\psi_j}$ remains unchanged. Equation (\ref{S3:eq:FormalAlgo}) can be rewritten as:
\begin{align}
& \dfrac{1}{\sqrt{2^m}} \dsum_{k=0}^{2^m-1} e^{i 2 \pi k \phi_j } \ket{k} 
\label{S3:eq:IterOpEnd}
\end{align}
then applying the inverse of the QFT to the state described in equation (\ref{S3:eq:IterOpEnd}) we figure out the probability to measure an individual output $j$ over the ancillary register given by:

\begin{equation}\label{S3:eq:ProbabilityPj}
P_j = \left|  \ \dfrac{1}{2^m}\dsum_{k=0}^{2^m} e^{i 2 \pi k (\phi_j - \frac{j}{2^m}) } \right|^2
\end{equation}

In equation (\ref{S3:eq:ProbabilityPj}), $j/2^m$ may be considered as a good estimator of $\phi_j$.
Since certainty occurs when $P_j = 1$ then,
from equation (\ref{S3:eq:ProbabilityPj}), the algorithm becomes deterministic when$$\phi_j - j/2^m = 0$$is fulfilled; the so called {\em zero theoretical failure rate condition}. That situation takes place when $\phi_j$ is expressed as a finite binary expansion by means of $m$ bits.
\subsection{The case of $m=n$: a simple example}
As a first approach for working out the set of unitary operators satisfaying the {\em zero theoretical failure rate condition} we face up the case where the number of ancillary states $m$, and the number of input quantum states $n$, are equal, $m=n$. Under this condition $\phi_j$ is written as follows:
\begin{equation}\label{S3:eq:EstimationEigenFirst}
2^l \phi_j =2^l \frac{j}{2^n}= \dsum_{k=0}^{n-1} j_k 2^{k-n+l}\quad \text{with}\quad j_k \in \{0,1\}\quad \text{and} \quad l=n-1\,\dots\,0
\end{equation}
 It is worth it noticing that for any $j_k$ fullfilling $k-n+l \geq 0$ there will be no contribution to the kicked phase shift. Therefore, equation (\ref{S3:eq:EstimationEigenFirst}), can be rewritten as its {\em principal value} which is given by:  

\begin{equation}\label{S3:eq:EstimationEigenSimple}
2^l \phi_j = \dsum_{k=0}^{n-1-l} j_k 2^{k-n+l} 
\end{equation}
Now, imposing equation (\ref{S3:eq:EstimationEigenSimple}) to equation
(\ref{S3:eq:FormalAlgo}), the ancillary register becomes
the QFT of the eigenvector $\ket{\psi_j}=\ket{j_{n-1}j_{n-2}\cdots j_1 j_0}$:
\begin{align}
& \gap \Normqv \bigotimes_{l=n-1}^{0} \left( \ket{0} 
	+ e^{i 2 \pi (2^l \phi_j) } \ket{1}  \right)
\notag \\
& =  \Normqv \bigotimes_{l=n-1}^{0} \left( \ket{0} 
	+ e^{i 2 \pi (\sum_{k=0}^{n-1-l} j_k 2^{k-n+l}) } \ket{1}  \right)
\\
& = \Normqv \left( \ket{0} + e^{i 2 \pi (0.j_0) } \ket{1} \right) 
	\cdots \left( \ket{0} + e^{i 2 \pi (0.j_{n-1} \cdots j_0) } \ket{1} \right) 
\end{align}

\begin{figure}[h]
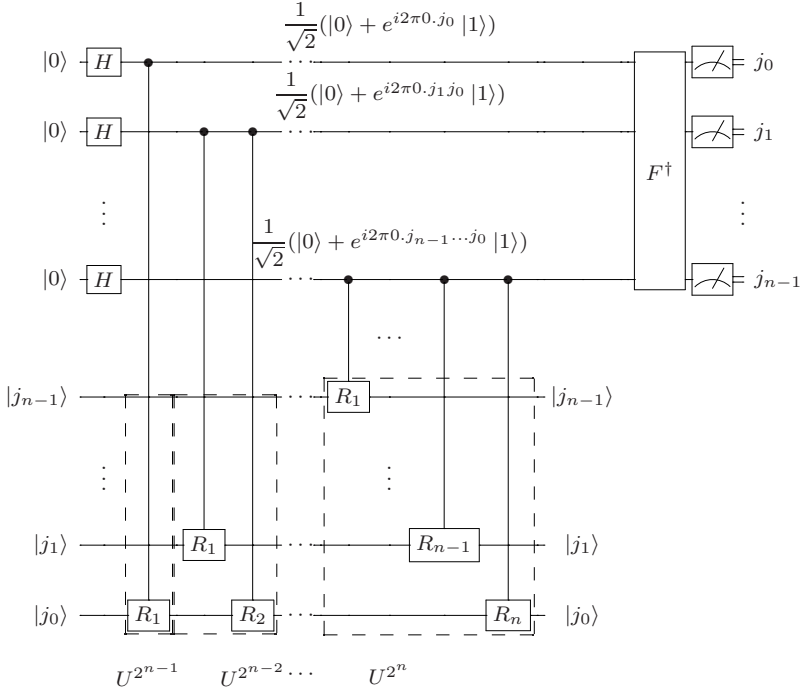

\include{2015_Figure04}
\caption{Detailed schema of a zero failure rate PEA using controlled
$z$--rotation gates}\label{S3:Fig:ZeroFailurePEA}
\end{figure}

In Figure \ref{S3:Fig:ZeroFailurePEA}, a PEA schema with a set $c\_U^{2^l}$ of 
operators fulfilling equation (\ref{S3:eq:EstimationEigenSimple}) is implemented in terms of a tensor composition of controlled $z$-rotation gates $R_k$ that act in the following way:$$R_k\ket{jm}=e^{i\frac{2 \pi(j\cdot m)}{2^k}}\ket{jm}$$So, $R_k$ performs a clockwise $\frac{(j\cdot m)}{2^{k-1}}\pi$ rotation only when the condition $j=m=1$ is fulfilled.

According to that, $U^{2^0}$ can be represented as a diagonal matrix whose 
eigenvectors are the elements of the standard basis and its corresponding 
eigenvalues (diagonal elements) are phase shifts. The particular case of 
$U^{2^0}$ for $n= 2$ is presented in Table \ref{S3:Tab:DetailedDescription}.
\renewcommand\multirowsetup{\centering}
\begin{table}[h]\centering
\begin{tabular}{ >{$}c<{$}  >{$}c<{$}  >{$}c<{$}  >{$}c<{$}  >{$}c<{$} }
\hline\noalign{\smallskip}
U^{2^0} & \ket{\psi} = \ket{j_1 j_0} & \lambda & \phi = j/2^n & j\\
\noalign{\smallskip}\hline\noalign{\smallskip}
\multirow{4}{*}{$
 \eins_{4\times 4}
\begin{bmatrix}
e^{i 2 \pi (0)}\\
e^{i 2 \pi (\frac{1}{4})}\\
e^{i 2 \pi (\frac{1}{2})}\\
e^{i 2 \pi (\frac{3}{4})}
\end{bmatrix}
$}
 & \ket{00} & 1 & 0 & 0 \\[1mm] 
 & \ket{01} & i & 1/4 & 1 \\[1mm]  
 & \ket{10} & -1 & 1/2 & 2 \\[1mm]  
 & \ket{11} & -i & 3/4 & 3\\
\hline\noalign{\smallskip}
\end{tabular}
\caption{Detailed description of $U^{2^0}$ for $n=2$ and $f_k \in 
\left\lbrace 0,1\right\rbrace $ }\label{S3:Tab:DetailedDescription}
\end{table}

\section{A deterministic dircuit for counting qubits: the Quantum Abacus}\label{S4:TQA}

The key point now is to implement a deterministic PEA based quantum circuit for counting how many qubits there are in a sequence. From a computational point of view this problem can be reduced to describe the circuit for counting how many 1's and 0's there are in the mentioned sequence given the current computational state of the qubit.  In fact, this circuit can be considered as the composition of two independent circuits, once for counting 1's and another one for counting 0's. Since both circuits are formally equivalents, just one will be considered for our research purposes. As it was seen in the previous section, the way to tackle this problem involves a specific definition of parameter $\phi_j$ in terms of the problem to be solved. If $\ket{q_{n-1}q_{n-2}\cdots q_1 q_0}$ is the input register containing a sequence of $n$ qubits, then the number of qubits with 1's, $N_{\ket{1}}$, or the number of qubits with 0's, $N_{\ket{0}}$, in this sequence is described in decimal form as:

\begin{align}\label{S4:eq:SequenceInDecimal}
N_{\ket{1}} & = \dsum_{k=0}^{n-1} q_k && \text{or} & 
	N_{\ket{0}} & = \dsum_{k=0}^{n-1}\bar{q}_k && \text{for} 
	& q_k \in \left\lbrace 0,1\right\rbrace
\end{align}
For the sake of simplicity and since both are formally identical, just one, the case 
when the qubit is $\ket{1}$, will be explicitly worked out. Then $\phi_{\ket{1}}$ is defined 
as:

\begin{align}\label{S4:eq:SequenceForOnes}
\phi_{\ket{1}} & = \dfrac{1}{2^m}N_{\ket{1}} \, = \, 
	\dfrac{1}{2^m}\dsum_{k=0}^{n-1} q_k
	\, = \, \dsum_{k=0}^{n-1} q_k 2^{-m} && \text{where}
	& q_k \in \left\lbrace 0,1\right\rbrace
\end{align}
where $m$ is the number of qubits of the ancillary register such that
$m = \lceil(\log_2 n)\rceil$. This condition is introduced in 
order to guarantee an optimal binary representation of the number of 1's after the 
measuring process. From the zero theoretical failure rate condition
$\phi_{\ket{1}}-j/2^m = 0$, so, the next identity is yielded:

\begin{align}\label{S4:eq:HowPEACircuit}
\dfrac{1}{2^m}\dsum_{k=0}^{n-1} q_k- \dfrac{1}{2^m}\dsum_{l=0}^{m-1} j_l 2^l
	& = 0 && \longrightarrow & \dsum_{k=0}^{n-1} q_k & = \dsum_{l=0}^{m-1} j_l 2^l
\end{align}

Equation (\ref{S4:eq:HowPEACircuit}) expresses how a PEA based circuit defined with a phase according to equation (\ref{S4:eq:SequenceForOnes}) will return, after applying the inverse of QFT, the binary representation of the number of 1's in the input sequence of qubits. Following the previous ideas it is not difficult to build the corresponding chain of transformations for a $U^{2^l}$ satisfying  
(\ref{S4:eq:SequenceForOnes}):

\begin{align}
\ket{0^{\otimes (m)}, q^{\otimes (n)}} & 
	\xrightarrow{H^{\otimes (m)} \otimes\hat{I}^{\otimes (n)}}
	\dfrac{1}{\sqrt{2 m}} \bigotimes_{l=m-1}^{0} \left( \ket{0} + \ket{1} \right)
	\ket{q^{\otimes (n)}}
\\
\xrightarrow{c\_U^{2^l}} & = \dfrac{1}{\sqrt{2 m}}
	 \bigotimes_{l=m-1}^{0} \left( \ket{0}\, + \ket{1}
	 e^{i 2 \pi 
	 \left( 2^l \sum_{k=0} ^{n-1} q_k 2^{-m} \right)} \right)
	 \ket{q^{\otimes (n)}}
\end{align}
where the ancillary register takes the form:
\begin{align}\label{complexity}
\gap & \bigotimes_{l=m-1}^{0}\ket{\varphi_l}\equiv\bigotimes_{l=m-1}^{0}\dfrac{1}{\sqrt{2}}
	\left( \ket{0} + e^{i 2 \pi 2^{l-m}\left(\sum_{k=0} ^{n-1} q_k \right)}
	\ket{1} \right)
\\ \nonumber
& = \dfrac{1}{\sqrt{2 m}} \bigotimes_{l=m-1}^{0}
	\left( \ket{0} + e^{i \pi \left(\sum_{k=0} ^{n-1} q_k \right)}
	\ket{1} \right)
	\cdots
	\left( \ket{0} + e^{i \pi \frac{1}{2^{m-1}} \left(\sum_{k=0} ^{n-1} q_k \right)}
	\ket{1} \right) \label{S4:eq:AncillaryRegister}
\end{align}

\begin{figure}[h]
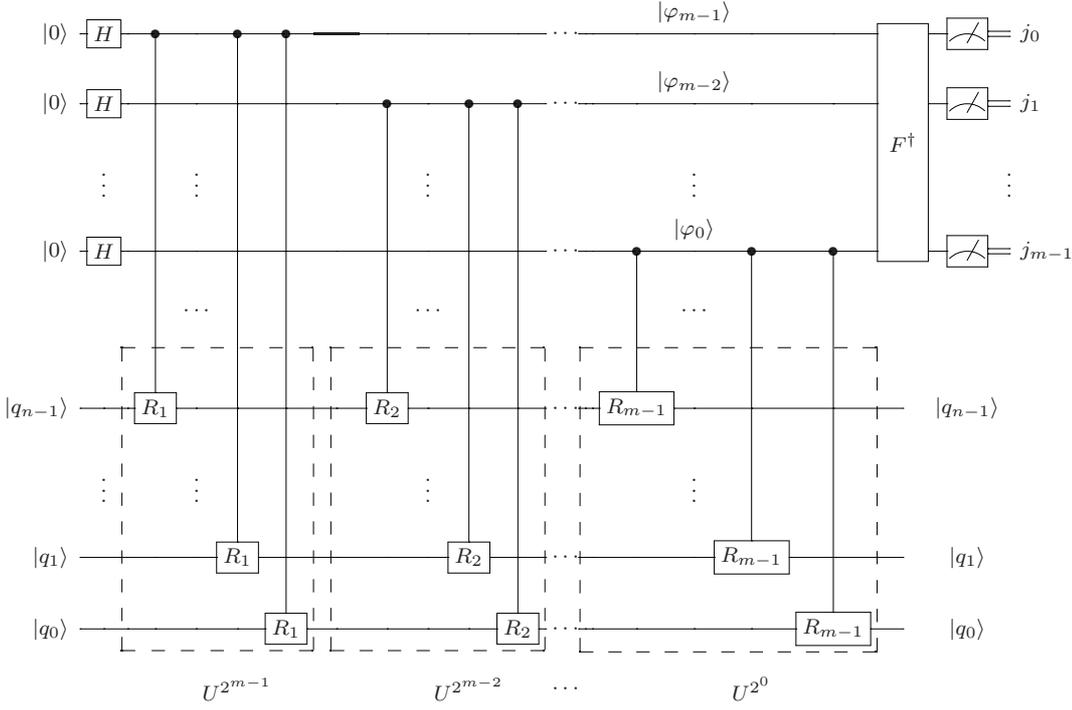

\include{2015_Figure05}
\caption{The deterministic circuit for counting how many 1's there are in a 
particular sequence of qubits}\label{S4:Fig:DeterministicCircuit}
\end{figure}

Again, a tensor composition of controlled $z$--rotation gates is involved to reproduce the corresponding phase shift addition. In order to yield a correct representation of counting register a set of Swap gates are placed after the inverse QFT.
The circuit corresponding to equation (\ref{complexity}) is scalable and its algorithmic complexity is easely worked out just taking into account the following issues:
\begin{itemize}
\item there is an arrangement of $\bigO (\log n)$ dashed boxes 
\item each dashed box contains exactly $n$ quantun rotational gates
\end{itemize}
This gives an amount of $\bigO (n \log n)$ quantum gates as it is shown in Figure \ref{S4:Fig:DeterministicCircuit} where, a deterministic quantum circuit for counting 1's from a register containing a sequence of $n$ qubits, is drawn.

In Table 2, the particular case of $U^{2^0}$ when $n=3$ for the previous circuit, is presented. 

\newcommand{\expi}[1]{e^{i 2 \pi (#1)}}
\begin{table}[h]\centering
\begin{center}
\begin{tabular}{>{$}c<{$}>{$}c<{$}>{$}c<{$}>{$}c<{$}>{$}c<{$}}
\hline\noalign{\smallskip}
U^{2^0} & \ket{q_2 q_1 q_0} & \lambda & \phi & q_{\ket{1}} = \dsum_{k=0}^{2}q_k = j \\
\noalign{\smallskip}\hline\noalign{\smallskip}
\multirow{8}{*}{$
 \eins_{8\times 8}
\begin{bmatrix}
\expi{0}\\ \expi{\frac{1}{4}} \\ \expi{\frac{1}{4}} \\ \expi{\frac{1}{2}} \\
\expi{\frac{1}{4}} \\ \expi{\frac{1}{2}} \\ \expi{\frac{1}{2}} 
\\ \expi{\frac{3}{2}}
\end{bmatrix} 
$}
 & \ket{000} & 1 & 0 & 0 \\[0.75mm]
 & \ket{001} & i & 1/4 & 1 \\[0.75mm]
 & \ket{010} & i & 1/4 & 1 \\[0.75mm]
 & \ket{011} & -1 & 1/2 & 2 \\[0.75mm]
 & \ket{100} & i & 1/4 & 1 \\[0.75mm]
 & \ket{101} & -1 & 1/2 & 2 \\[0.75mm]
 & \ket{110} & -1 & 1/2 & 2 \\[0.75mm]
 & \ket{111} & -i & 3/2 & 3 \\[0.75mm]
 \hline\noalign{\smallskip}
\end{tabular}
\end{center}
\caption{Detailed description of $U^{2^0}$ for $n=3$ and $q_k \in 
\left\lbrace 0,1\right\rbrace $ }\label{S4:Tab:DetailedDescription}
\end{table}

\begin{figure}[!h]
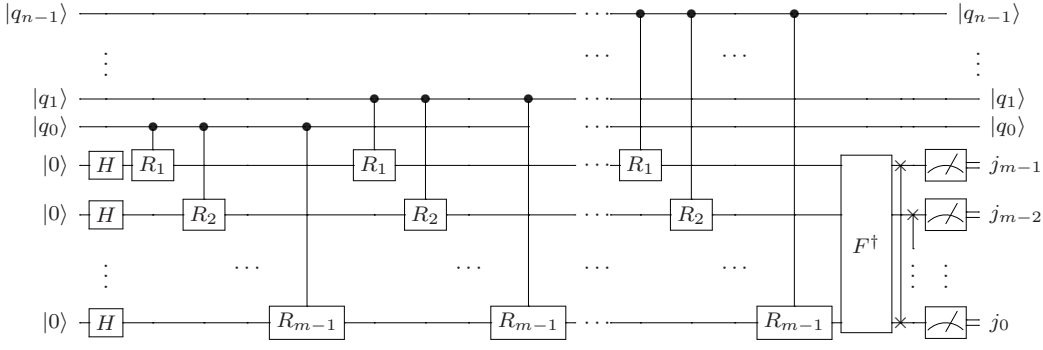

\include{2015_Figure08}
\caption{An equivalent circuit for counting 1's}
\label{S4:Fig:DeterministicOnes}
\end{figure}
\newcommand{\funktiong}{
\ensuremath{
\dfrac{1}{\sqrt{2}} (\ket{0} + e^{i 2 \pi\; 0.j_0} \ket{1} )
}}
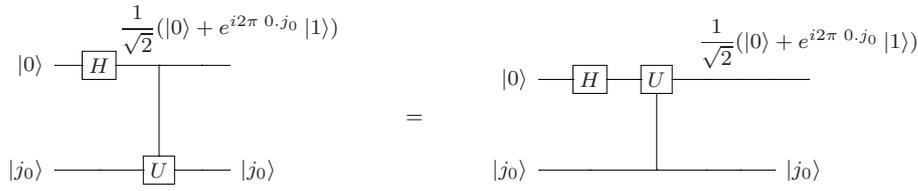
\begin{figure}[!h]
\begin{center}
\vspace{1cm}
\begin{tabular}{lcr}
\parbox{\textwidth}{%
\Qcircuit @C=1.2em @R=1.7em {%
\lstick{\ket{0}} & \gate{H} & \qw \qwx[2] & \qw
& \ustick{\funktiong} \qw
\\
& & &
\\
\lstick{\ket{j_0}} & \qw & \gate{U} & \qw 
& \rstick{\ket{j_0}} \qw
} 
} 
& \hspace{1.8cm}  = \hspace{1cm} &
\parbox{\textwidth}{%
\Qcircuit @C=1.5em @R=1.7em {%
\lstick{\ket{0}} & \gate{H} & \gate{U} & \qw & \qw
& \qw &\ustick{\funktiong} \qw
\\
& & &
\\
\lstick{\ket{j_0}} & \qw & \qw \qwx[-2] & \qw  & \qw
&\rstick{\ket{j_0}} \qw
} 
} 
\end{tabular}
\end{center}

\caption{Ambiguity of who controls whom}
\label{S4:Fig:Ambiguity}
\end{figure}
Given the ambiguity of who controls whom that occurs in those quantum circuits where controlled--$U$ gates are involved (Figure \ref{S4:Fig:Ambiguity}), an equivalent circuit for counting 1's may be proposed (Figure \ref{S4:Fig:DeterministicOnes}) reorganizing the circuit in Figure \ref{S4:Fig:DeterministicCircuit}. In this case the controlled $z$--Rotation gates are applied directly on the ancillary qubits.

This arrangement presents interesting similitudes with a classical abacus device where controlled shifts, applied on the ancillary qubits, play the role of tokens for counting. 
As it will be seen in the next section, this special feature  becomes the base of an encoding system for integer numbers.

\section{ Building the framework for quantum encoding}\label{S5:TQACodification}
The previous schema, without the control qubits, may be used for building the framework for quantum encoding since, any binary number, may be encoded in terms of shifts.
The unitary operator associated to this circuit will be denoted, from now on, as $U^c$. The set of transformations for this encoding process over a $n$ qubits quantum register is written as follows:

\begin{align}
& \gap \ket{0^{\otimes (n)}} \xrightarrow{H^{\otimes (n)}} 	
	\Normqv \bigotimes_{l=n-1}^{0} \left( \ket{0} + \ket{1} \right)
	 \xrightarrow{U^{c^{\otimes (n)}}} \\
 & = \Normqv \bigotimes_{l=n-1}^{0} 
	 \left( \ket{0} + e^{i 2 \pi \phi_d(l)} \ket{1} \right)
\end{align}
From equation (\ref{S4:eq:AncillaryRegister}), the encoding process is carried out through the phase $2 \pi \phi_d(l)$ where the factor $\phi_d(l)$\footnote{From now on, and for sake of simplicity, we will refer to the factor $\phi_d(l)$ as the {\it phase} $\phi_d(l)$.} is given by:

\begin{equation}\label{S5:eq:PhasePhi_l}
\phi_d(l) = \left( d \mod 2^{n-l} \right) \dfrac{1}{2^{n-l}}
\end{equation}
In equation (\ref{S5:eq:PhasePhi_l}), $d$ is the decimal representation of the encoded data. Obviously, $\phi_l$ can be rewritten as:

\begin{gather}\label{S5:eq:PhasePhi_l_rew}
\phi_d(l) = \left( d \mod 2^{n-l} \right) \dfrac{1}{2^{n-l}}
	= \dsum_{k=0}^{n-l-1} d_k 2^{k-n+l}
	= \left( \dfrac{d_{n-l-1}}{2} + \dfrac{d_{n-l-2}}{4} + \cdots
	+ \dfrac{d_0}{2^l} \right) 
\end{gather}
In equation (\ref{S5:eq:PhasePhi_l_rew}), $\phi_d(l)$ is the phase yielded as a result of applying the QFT to a register containing the binary representation of $d$. Then, if the inverse of QFT, followed by a set of Swap gates, are applied over the register the binary representation of the encoded data is retrieval.

\begin{gather}
S w^{\otimes (n)} \cdot F^{\dagger \otimes (n)}
	\left( \Normqv \bigotimes_{l=n-1}^{0} 
	\left( \ket{0} + e^{i 2 \pi \phi_d(l)} \ket{1} \right)
	\right) = \ket{d_{n-1} \cdots d_0}
\end{gather}

\begin{figure}[h]
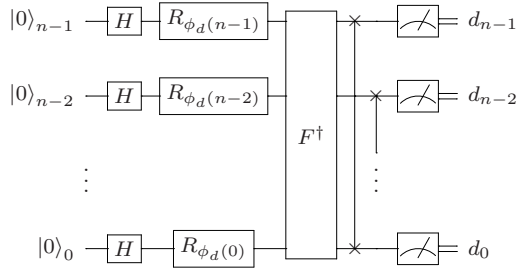

\include{2015_Figure09}
\caption{a) General quantum encoding circuit where $R_{\phi_d(l)}$ is the quantum gate that performs the phase rotation of the angle $2 \pi \phi_d(l)$.}
\label{S5:Fig:QCGeneral}
\end{figure}

\begin{figure}[h]
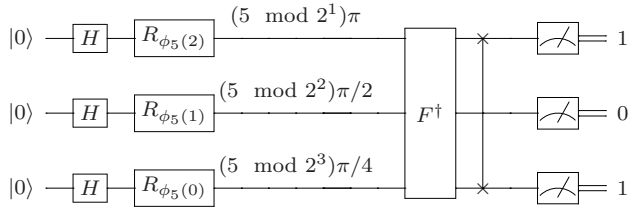

\include{2015_Figure10}
\caption{b) Codification of number 5 and the retrieval process to archieve its binary representation.}
\label{S5:Fig:QCNumberFive}
\end{figure}

The general quantum encoding circuit (QC) and an instance representing number 5 are shown in Figures \ref{S5:Fig:QCGeneral} and \ref{S5:Fig:QCNumberFive}. 

\newcommand{\funktionh}[2]{%
\ensuremath{%
(d \mod #1) \frac{\pi}{2^{#2}}
}}

\newcommand{\funktionk}[2]{%
\ensuremath{%
(5 \mod #1) \pi #2
}}

\section{Implementation of a Quantum Array using the quantum encoding Circuit}\label{S6:TQAImplemantation}

\begin{figure}[h!]
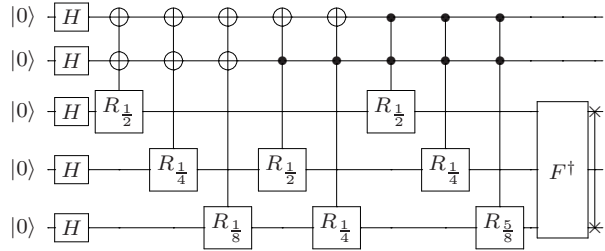

\include{2015_Figure11}
\caption{a) the corresponding one storing the numbers 1,2,0,5 according to this order $(\ket{0, 1} + \ket{1, 2}+ \ket{2, 0} + \ket{3, 5})$.}
\label{S6:Fig:StoringNumbers}
\end{figure}

\begin{figure}[h!]
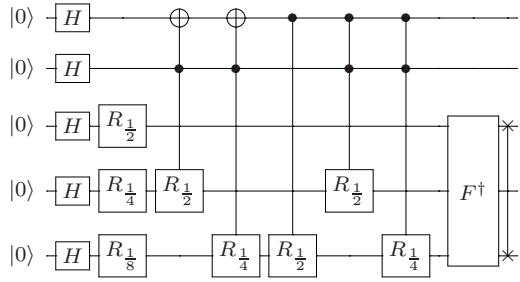

\include{2015_Figure12}
\caption{b) the corresponding one storing the first four odd numbers 
$(\ket{0, 1} + \ket{1, 3}+ \ket{2, 5} + \ket{3, 7})$.}
\label{S6:Fig:StoringOdd}
\end{figure}

An application of this encoding system is related to the implementation of a quantum data structure that we called as a {\em quantum array}. From a classical point of view an array, also known as vector, is a data structure where stored data of a specific type is accessible through an index.  This description can be used in the quantum context defining the corresponding storage structure and its fundamental operations. In the quantum world, a collection of data may be stored in a single register through an evenly distributed superposition of states. This way of storage comprises $2^m$ classical states on just one $m$ qubit register. The quantum array can be implemented from a superposition of states dividing the register in two different logical parts. The first one storages the most significant qubits and denotes the index that will be used for accessing to the data while, the second one, the logical part with the less significant ones, makes reference to the data itself. The number of qubits of each part will depend on the number of elements of the database and the size of the data respectively. For the sake of simplicity, the number of elements of the array will be always a power of two. In practice, this is not a problem since the array can be padded with zeros. So, if the quantum array size is $M=2^m$ then the index will be stood for through $m$ qubits in such a way that, if $p$ is the necessary number of qubits for a suitable representation of data, then the quantum register structure takes this form:
\begin{equation}
\ket{j_{m-1} \cdots j_0, d_{p-1} \cdots d_0}
\end{equation}
Given this basic structure, the creating and updating operations are defined as follows\footnote{The operation for data retrieving will be carried out by means of the generalization of Grover's algorithm \cite{Brassard2002}.}:
\paragraph{Creating a Quantum Array}
The following steps are involved for creating a specific quantum array:
\begin{enumerate}
\item An initial ancillary register of $m+p$ qubits is set to 0 and a set of $m+p$ Hadamard gates are applied on each ancillary qubit in order to generate the initial superposition.
\item Next, the corresponding $U^c$ operator is applied to encode the data.
\item Finally, the application of the inverse of QFT on this latter qubits to retrieve the binary representation of data.
\end{enumerate}
It is important to remark that $U^c$ operator can be implemented combining both, multiple controlled shift gates and just unary shift gates. In the former case, the control action is achieved through the index qubits in such a way that each superposition state generated on the former $m$ qubits is linked to a specific phase shift encoding, in this manner:

\newcommand{\Normqvd}[2][1]{\dfrac{#1}{\sqrt{2^#2}}}
\begin{align}
& \gap \ket{0^{\otimes (m)}, 0^{\otimes (p)}} \xrightarrow{H^{\otimes (m+p)}}
	\dfrac{1}{\sqrt{2^m}} \dsum_{j=0}^{2^m-1} \ket{j} \dfrac{1}{\sqrt{2^p}}
	\bigotimes_{l=1}^{p} \left( \ket{0} + \ket{1} \right)
	\xrightarrow{U^c}
\\
& = \Normqvd{m} \dsum_{j=0}^{2^m-1} \ket{j} U^c 
	\left(\dfrac{1}{\sqrt{2^p}} \bigotimes_{l=1}^{p}
	\left( \ket{0} + \ket{1} \right) \right) 
\\
& = \Normqvd{m} \dsum_{j=0}^{2^m-1} \ket{j}
	\left(\dfrac{1}{\sqrt{2^p}} \bigotimes_{l=1}^{p}
	\left( \ket{0} + e^{i 2 \pi (\phi^j+ \phi)_l} \ket{1} \right) \right)
	\quad \xrightarrow{F_{p}^{\dag}}
\\
& = \Normqvd{m} \dsum_{j=0}^{2^m-1} \ket{j}
	\hat{F}_p^{\dag} \left( \dfrac{1}{\sqrt{2^p}} \bigotimes_{l=1}^{p}
	\left( \ket{0} + e^{i 2 \pi (\phi^j+ \phi)_l} \ket{1} \right) \right)
\\
& = \Normqvd{m} \dsum_{j=0}^{2^m-1} \ket{j,d^j}
\end{align}
where $d^j$ is the decimal representation of data stored in the $j$ position of the quantum array. The encoding process is carried out through the corresponding phase shift $(\phi^j+ \phi)_l$ such that $\phi^j$, dependent shift phase of $j$, is achieved through the multiple controlled phase shift gates and the parameter $\phi$\footnote{A parameter representing a shift phase independent of $j$. }, wether elementary phase shift gates are involved. Both phases are 
expressed through equation (\ref{S5:eq:PhasePhi_l}). In Figures 
\ref{S6:Fig:StoringNumbers} and \ref{S6:Fig:StoringOdd}, two schematic circuits summarizing the previous description for storing numbers are presented: 
\begin{itemize}
\item[a)] The first one stores any four numbers that may represent, for instance, a contact list. 
\item[b)] The second one stores the first four odd numbers.
\end{itemize}

\begin{figure}[h!]
\begin{center}
\begin{tabular}{lcr}
\parbox{3cm}{%
\Qcircuit @C=0.6em @R=1.0em {%
\lstick{m-1}
& \qw & \qw & \qw & \qw & \qw
& \qw & \qw & \qw & \qw & \qw & \qw
\\
\lstick{m-2}
& \qw & \qw & \qw & \qw & \qw
& \qw & \qw & \qw & \qw & \qw & \qw
\\
&  \push{\vdots} &
& & & & &
&  \push{\vdots} & & &
\\
\lstick{1}
& \qw & \qw & \qw & \qw & \qw
& \qw & \qw & \qw & \qw & \qw & \qw
\\
\lstick{0}
& \qw & \qw & \qw & \qw & \ctrlo{1} \qwx[2]
& \qw & \qw & \qw & \qw & \qw & \qw
\\
&  &  & 
&  & \push{\hspace{6pt}/ p} &  &  & 
&  & & &
\\
\lstick{p-1} & \multigate{4}{F} & \qswap \qwx[4]
& \qw & \qw & \multigate{4}{1}
& \multigate{4}{F^{\dag}} & \qswap \qwx[4] & \qw & \qw &  \qw & \qw
\\
\lstick{p-2} & \ghost{F} & \qw
& \qw & \qswap \qwx[2] & \ghost{1}
& \ghost{F^{\dag}} & \qw & \qw & \qswap \qwx[2] &  \qw & \qw
\\
\push{\vdots} & &
& \push{\cdots} & &
&  & & \push{\cdots} & & \push{\vdots} &
\\
\lstick{1} & \ghost{F} & \qw
& \qw & \qswap & \ghost{1}
& \ghost{F^{\dag}} & \qw & \qw & \qswap &  \qw & \qw
\\
\lstick{0} & \ghost{F} & \qswap
& \qw & \qw & \ghost{1}
& \ghost{F^{\dag}} & \qswap & \qw & \qw &  \qw & \qw
} 
} 
\end{tabular}
\end{center}

\caption{Adding 1 just over the even positions of a predefined quantum array.}
\label{S6:Fig:AddingOneQArray}
\end{figure}
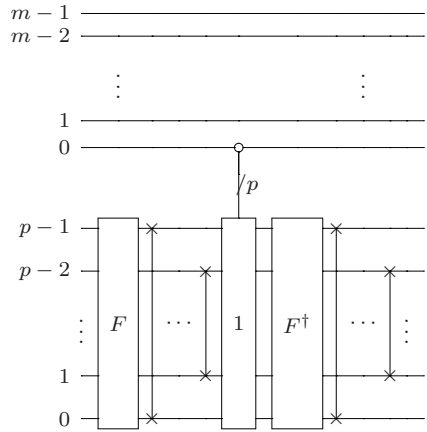

It is important to realize that, when the information to be stored is described by an arithmetic series (latter example), the quantum array can be implemented taking advantage of quantum parallelism.
\paragraph{Updating a Quantum Array.}
Once the quantum array is created any element within it may be modified.  In fact, through the encoding system, a subset of the quantum array states or all of them may be modified simultaneously. In Figure \ref{S6:Fig:AddingOneQArray}\footnote{In Figure \ref{S6:Fig:AddingOneQArray} the following equivalences have been applicated:
\begin{center}
\begin{tabular}{lcr}
\begin{tabular}{lcr}
\parbox{\textwidth}{
\Qcircuit @C=0.3em @R=0.3em {%
& \ctrl{3} & \ctrl{4} & \qw & \ctrl{5} & \qw
\\
& & & \push{\cdots} & &
\\
& & &  & &
\\
& \gate{R_{\frac{1}{2}}} & \qw & \qw & \qw & \qw
\\
& \qw & \gate{R_{\frac{1}{4}}} & \qw & \qw & \qw
\\
& \qw & \qw & \qw & \gate{R_{\frac{1}{2^m}}} & \qw
\gategroup{3}{1}{6}{5}{0.3em}{--}
} 
} 
& = &
\parbox{\textwidth}{
\Qcircuit @C=0.3em @R=0.7em {%
& \qw & \ctrl{3} & \qw
\\
& & {\;\;\; / m} &
\\
& & &
\\
& \qw & \multigate{2}{1} & \qw
\\
& \qw & \ghost{1}  & \qw
\\
& \qw & \ghost{1}  & \qw
} 
} 
\end{tabular}
& \text{and} &
\begin{tabular}{lcr}
\parbox{1cm}{
\begin{equation*}
\Qcircuit @C=0.3em @R=0.5em {%
& \qw & \ctrlo{1} & \qw &\qw
\\
& & \push{\vdots}
} 
\end{equation*}
} 
 = & 
\parbox{0.5cm}{
\begin{equation*}
\Qcircuit @C=0.3em @R=0.5em {%
& \gate{X} & \ctrl{1} & \gate{X} & \qw
\\
& & \push{\vdots} & &
} 
\end{equation*}
} 
\end{tabular}
\end{tabular}
\end{center}
}, a quantum circuit for adding a number, for instance 1, just over the even positions of a predefined quantum array, is described. This operation is performed simultaneously over all the involved states. 
In order to clarify this process a detailed mathematical description step by step is shown:
\begin{enumerate}
\item First a quantum register is prepared in a quantum superposition state $\ket{\psi}$ $$\ket{\psi} = \Normqvd{m} \dsum_{j=0}^{2^m-1} \ket{j, d^j}$$
\item The following quantum operators are applied$$\left( I^{\otimes m}\otimes (S w^{\otimes (p)} \cdot F^{\otimes (p)}) \right) \ket{\psi}$$
\item Letting the quantum register in the quantum state written below $\ket{\psi}$ $$\Normqvd{m} \dsum_{j=0}^{2^m-1} \ket{j} \bigotimes
	\left( \Normqvd{p} \bigotimes_{l=p-1}^{0} \left(  \ket{0} + e^{i 2 \pi d 2^{-p + l}} \ket{1} \right) \right)$$
\item Then the quantum operator $C\_1_{j\; \text{even}}$ is applied and an {\it even} and {\it odd} decomposition is achieved
\begin{align}
& \Normqvd{m} \dsum_{j\; \text{even} =0}^{2^m-1} \ket{j} \bigotimes
	\left( 
	\Normqvd{p} \bigotimes_{l=p-1}^{0} \left(
	 \ket{0} + e^{i 2 \pi (d^j+ 1) 2^{-p + l}} \ket{1} \right)
	\right) \notag
\\ 
& \gap[=] + \Normqvd{m} \dsum_{j\; \text{odd} =0}^{2^m-1} \ket{j} \bigotimes
	\left( 
	\Normqvd{p} \bigotimes_{l=p-1}^{0} \left(
	 \ket{0} + e^{i 2 \pi d^j 2^{-p + l}} \ket{1} \right)
	\right)
\end{align}
\item Finally the QFT inverse acts$$I^{\otimes m}\otimes (S w^{\otimes (p)} \cdot F^{\otimes (p)})$$and the quantum register takes the following form$$ \ket{\psi'}\; = \; \Normqvd{m} \dsum_{j=0}^{2^m-1} \ket{j, d'^j}$$where $d'^j = d^j + 1$, if $j$ is an {\it even} number and $d'^j = d^j$  if $j$ is an {\it odd} number.
\end{enumerate}

\section{Conclusions}\label{S7:Conclusions}

Our contribution on this paper has been to provide a quantum framework for manipulating data on a quantum data structure. This framework, the Quantum Abacus based enconding system, allows to work with any integer numbers in terms of phase shifts instead of its corresponding binary representation. In this context, a Quantum Abacus is a circuit for counting qubits depending on its current computational state. A formal description of this latter circuit, in terms of a QFT based PEA schema, is also presented. In order to show the possibilities of this framework, the implementation of creating and updating operations of a quantum data structure (Quantum Array) has been achieved, taking advantage, when it has been possible, of quantum parallelism.

\begin{acknowledgements}
This work was partially supported by the Spanish Ministerio de Econom\'ia y Competitividad
(MINECO) under project MTM2014-57129-C2-1-P.
\end{acknowledgements}




\end{document}